\documentclass[letterpaper]{article} 
\usepackage[]{aaai25}  
\usepackage{times}  
\usepackage{helvet}  
\usepackage{courier}  
\usepackage[hyphens]{url}  
\usepackage{graphicx} 
\usepackage{natbib}  
\usepackage{caption} 
\usepackage{algorithm}
\usepackage{algorithmic}

\usepackage{newfloat}
\usepackage{listings}
\DeclareCaptionStyle{ruled}{labelfont=normalfont,labelsep=colon,strut=off} 
\floatstyle{ruled}
\newfloat{listing}{tb}{lst}{}
\floatname{listing}{Listing}
\usepackage{latexsym}
\usepackage{url}
\usepackage[T1]{fontenc}
\usepackage{subcaption}
\usepackage{soul}
\usepackage{array} 
\usepackage{amsmath}
\usepackage{amssymb}

\usepackage{xcolor}
\newcommand{\answerYes}[1]{\textcolor{blue}{#1}} 
\newcommand{\answerNo}[1]{\textcolor{teal}{#1}} 
\newcommand{\answerNA}[1]{\textcolor{gray}{#1}}

\title{Jointly modelling the evolution of social structure \\and language in online communities}
\author{
    Christine de Kock
}
\affiliations{
    University of Melbourne\\
    School of Computing and Information Systems\\
    christine.dekock@unimelb.edu.au

}

\usepackage{bibentry}

\begin{document}
\nocopyright
\maketitle

\begin{abstract}
Group interactions take place within a particular socio-temporal context, which should be taken into account when modelling interactions in online communities. We propose a method for jointly modelling community structure and language over time. Our system produces dynamic word and user representations that can be used to cluster users, investigate thematic interests of groups, and predict group membership. We apply and evaluate our method in the context of a set of misogynistic extremist groups. Our results indicate that this approach outperforms prior models which lacked one of these components (i.e.\ not incorporating social structure, or using static word embeddings) when evaluated on clustering and embedding prediction tasks. Our method further enables novel types of analyses on online groups, including tracing their response to temporal events and quantifying their propensity for using violent language, which is of particular importance in the context of extremist groups. 
\end{abstract}

\section{Introduction}\label{sec:intro}
Online communities present a unique opportunity for studying the evolution of groups. Two aspects are often considered when researching online communities: the structure of the community (\textit{who talks to whom}; e.g. \citealp{gialampoukidis2017terrorism}) and the language they use (\textit{what do they say}; e.g. \citealp{danescu2013no}). While most existing works tend to explore these facets in isolation, the premise of our work is that language and social structure are mutually informative expressions of group belonging that should be modelled jointly. Moreover, such interactions take place in a particular temporal context. On the internet, language changes at rapid rates \citep{stewart2018making} and the community itself morphs and changes focus in response to current events \citep{baele2023diachronic,ribeiro2021evolution}. 

To model the interaction of these dimensions, we propose a shared matrix factorisation model which jointly encodes linguistic and social evolution over time, yielding dynamic user and word embeddings. We evaluate this model in the context of misogynistic extremist groups (collectively known as \textit{the manosphere}) and illustrate the usefulness of the resulting embeddings in researching communities. 

The proposed model is an amalgamation of two prior works: \citet{appel2019temporally} developed a shared matrix factorisation approach which captures the temporal evolution of users, but modelled language as being static; while \citet{yao2018dynamic} used a similar method to construct dynamic word representations, but did not incorporate social information. By combining these two approaches, we obtain representations of users and words that evolve over time and are informed by their social context. This enables us to reason about the temporal evolution of users and their language in a shared space; a promising step to advance computational modelling of extremist groups. That said, this method is not specific to extremist groups and can be applied within any community for which social and linguistic information is available over multiple timesteps.

Experimental results (Section \ref{sec:results}) illustrate that our method improves over the abovementioned methods in several evaluation settings. Firstly, we cluster the dynamic user embeddings and find that our model obtains a statistically significant improvement in cluster purity when evaluated against subreddit labels. Secondly, we illustrate that the user embeddings can be used to predict future behaviour of an individual more accurately. In particular, we forecast embeddings and predict the subreddits with which a user will engage beyond the training window, finding that our model outperforms prior approaches in three different test formulations. 

Our method further enables novel types of analyses and visualisations of online communities, individuals within these groups, and their evolving interests and language. We provide a qualitative illustration of the importance of dynamic, temporally-grounded word embeddings in Section~\ref{sec:word_evolution} by investigating current affairs with a temporal dimension; e.g. the \textit{MeToo} movement. Finally, we present a novel characterisation of violent language in subgroups within the manosphere, which leverages the shared user and word embedding space (Section \ref{sec:splintering}). Our results show that Incel-related groups have the highest propensity for violent language among the manosphere subgroups, and that there is evidence of smaller, more violent groups, characteristic of the ``splintering'' phenomenon in extremist groups. 

\section{Modelling users and words over time}\label{sec:background}
The goal of this work is to construct dynamic word and user embeddings in a shared space by accounting for three factors: language, social dynamics, and time. Such representations can form the basis for analysing and predicting the behaviour of individuals and groups, which is of particular importance within extremist communities. In this section, we discuss the relationships between each pair of factors as explored in prior work, and highlight their significance in the context of extremist groups.

\subsection{Time and groups} 
Online communities are highly dynamic, with continuous developments at the micro level (through users entering and leaving the group) as well as the macro level (through evolving sub-community structures). Group polarisation theory \citep{myers1976group, sunstein2002law} provides one explanation for this phenomenon, hypothesising that discussions within homogeneous groups intensify extremist positions. This intensification forces results in the departure of moderate voices, creating smaller and increasingly radical communities. This can result in ``splintering'', a term used in counter\-terrorism research for the process whereby extremist ideologies tend to fragment over time into a range of sub-ideologies supported by rival factions, with minority splinter groups going towards increasing radicalism \citep{baele2023diachronic}. 

Fracturing in groups can also be the result of external pressures, such as deplatforming. \citet{di2024characterizing} map nine mainstream and alt-tech sites and demonstrate that user migration driven by moderation pressure produces ideologically homogeneous ``echo-platforms'' characterised by a greater prevalence of unreliable content and a heightened ideological uniformity. Similarly, \citet{vu2024no} study a deplatforming attempt on a doxxing platform, finding that it primarily affected casual users rather than core members, while also attracting new users who exhibit higher levels of toxicity at the time of joining. 

At the micro level, individuals also evolve over time in their level of dedication to a community. While the definition of radicalisation is widely disputed by scholars, it is generally agreed that it occurs through a gradual process over time \citep{silke2010psychology,della2012guest}. 

\subsection{Groups and language}
Language has been shown to be strongly tied to social connectedness in online groups, as it is used to signal in-group status \citep{drake1980social}. Social identity theory \citep{tajfel1979integrative} provides a basis for this dynamic, positing that individuals derive self-concept from their group memberships, leading them to accentuate distinctions between their in-group and relevant out-groups. Contemporary research in sociolinguistics has observed these dynamics in the context of online communities. For example, \citet{zhang2018conversations} show that communities with more specialised and dynamic language are more likely to retain their users. Longitudinal tracking of user lifecycles further shows that users linguistically align themselves with a community as they become embedded in it \citep{danescu2013no}. This linguistic alignment wanes before departure, making shifting language a dynamic proxy for changing identification with the group.

In the context of extremism, the use of group-specific neologisms has been proposed as a proxy measure for the amount of radicalisation influence an individual is under (e.g. \citealp{fernandez2018understanding}). \citet{de2024investigating} found that the use of group jargon is highly correlated with post volume and social connectedness in extremist groups, and that early adoption of group language is predictive of eventual levels of radicalisation indicators. 

\subsection{Language and time} 
The development of language over time is well-established within NLP research \citep{wang2020static}. The internet, having a strong social component and reduced impetus for formality, is a particularly fertile domain for creating new words \citep{stewart2018making} and for imbuing existing words with new meanings \citep{mendelsohn2023dogwhistles}. \citet{bogetic2023race} states that the manosphere are prolific lexical innovators, creating new terminology at more rapid rates than other extremist groups. Per illustration, the manosphere lexicon of \citet{farrell2019exploring} includes more than 36 hateful neologisms for women. The rapid development of new language creates challenges in moderating toxic content online, since it can be difficult for outsiders to interpret \citep{mendelsohn2023dogwhistles}.

\subsection{Our contribution: combining all factors}
Our proposed model, detailed in Section \ref{sec:model}, learns dynamic user and word representations in the same space, informed by both linguistic and social context. To our knowledge, we are the first to model these factors jointly in the context of online communities. Comparing to systems that use static word representations or neglect social information, we obtain a statistically significant improvement in three evaluation settings, indicating that it is important to account for all three abovementioned interactions. Representing a community in this way allows for novel types of analyses of extremist groups, as explored in Sections~\ref{sec:user_evolution}, \ref{sec:word_evolution} and \ref{sec:splintering}.  Since this method relies on matrix factorisation, a well-established unsupervised modelling methodology, it is computationally efficient, interpretable and easily extendible.

\section{Model definition}\label{sec:model}
\begin{figure*}
    \centering
    \includegraphics[width=0.9\linewidth]{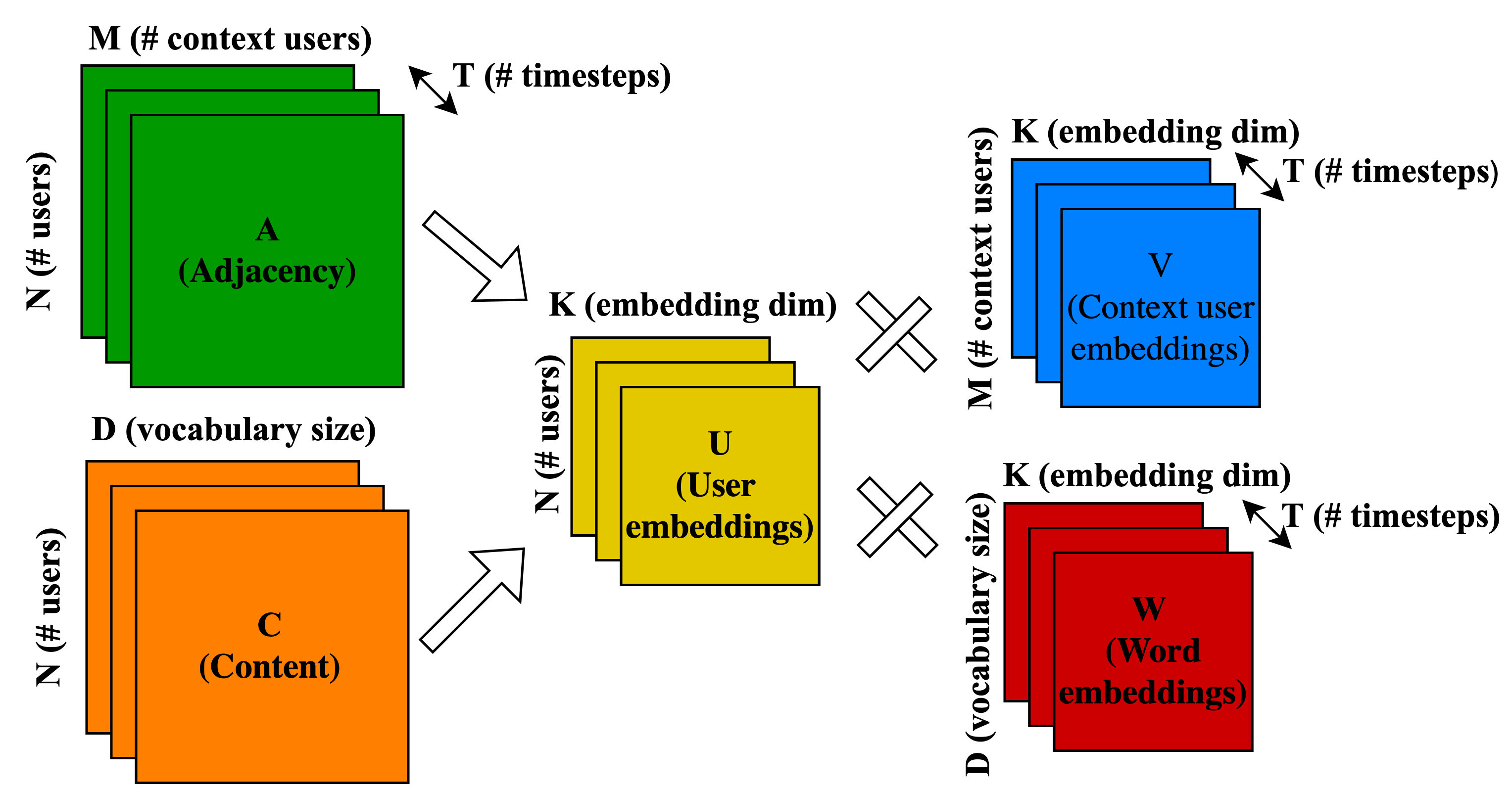}
    \caption{The proposed model architecture.}
    \label{fig:model}
\end{figure*}
Our proposed model jointly models social structure and language by decomposing a \textbf{social adjacency matrix} and a \textbf{language content matrix} over multiple timesteps. The model is illustrated in Figure \ref{fig:model}. In this section, we discuss its components, optimisations, and comparisons with other formulations.

\subsection{Source matrices}\label{sec:model.source}
The \textbf{social adjacency matrix} for a population of size $m$ is a sparse matrix defined as $A \in \mathbb{R}^{m \times n}$, where each row represents the social connections of a specific user to a set of $n$ other users, often referred to as \textit{context users}.
Each element $a_{ij}$ is calculated as the number of discussion threads that both user $i$ and user $j$ engaged in, with row-wise L1 normalisation.
Formally:
\begin{equation}\label{eq:adjacency}
a_{ij} = \frac{\vert \theta_i \cap \theta_j \vert}{\sum_{y=0}^{n} \vert \theta_i \cap \theta_y \vert}
\end{equation}
where $\theta_i$ is the set of discussion threads user $i$ has engaged in.
Intuitively, this formulation considers users to be similar if they interact with similar people. Similarly, given a vocabulary of size $d$, we construct a \textbf{language content matrix} $C\in\mathbb{R}^{m \times d}$ such that each element in $C$ represents the usage of a word $z$ by a user $i$, defined by its pointwise mutual information (PMI; \citealp{jurafsky2000speech}):
\begin{equation}
    \text{PMI}(z,i) = log(\frac{P(z|i)}{P(z)}),
\end{equation}
where $P(z)$ is determined by the frequency of $z$ in a background corpus.
We use the positive PMI (PPMI), which truncates scores at 0 to avoid overemphasising low-frequency events.
Intuitively, this formulation considers users to be similar if they have a proclivity for using the same uncommon words.

\subsection{Matrix factorisation}\label{sec:model.joint}
We use matrix factorisation to jointly decompose the social adjacency matrix $A$ into two matrices $U\in \mathbb{R}^{m \times k}$ and $V\in \mathbb{R}^{n \times k}$ such that $A \approx UV^T$. Concurrently, the language content matrix $C$ is decomposed into two matrices $U\in \mathbb{R}^{m \times k}$ and $W\in \mathbb{R}^{d \times k}$ such that $C \approx UW^T$. Importantly, $U$ is shared between both operations. The latent factor matrices ($U$, $V$, and $W$) are dense, lower-dimensional representations of users, context users and words, respectively. The decomposition of a matrix can be learned via a gradient-based optimisation algorithm with an objective function based on the Frobenius norm of the reconstruction error: $J = ||A - UV^T||^2$ (for the social adjacency matrix, or the equivalent for the content matrix). 

When decomposed independently based on $A$, the resulting matrix $U$ consists of an embedding for each of the $m$ users, which will be similar for users who often interact with the same set of context users. Similarly, when decomposed independently based on $C$, the resulting matrix $U$ consists of an embedding for each of the $m$ users, which will be similar for users who often use the same uncommon words. To learn user embeddings that are informed by both linguistic and social behaviours, we jointly decompose $A$ and $C$ such that the user embeddings $U$ need to be able to reconstruct both source matrices.

To allow for temporal evolution over $T$ time\-steps, we learn time-based latent factors $U_t$, $V_t$ and $W_t$ for $t \in \{1,...,T\}$ based on $A_t$ and $C_t$. To encourage alignment between embeddings over different timesteps, we follow \citet{appel2019temporally} and \citet{yao2018dynamic} in adding a temporal smoothing term (shown in Eq.~\ref{eq:objective}) that discourages large variations across timesteps. 

The final learning objective is given by:
\begin{equation}
\begin{split}
    J &= \sum_{t=1}^{T}
    ||A_t - U_tV_t^T||^2 +  ||C_t - U_tW_t^T||^2 \\
     &+ \sum_{t=1}^{T}\lambda_1(||V_t||^2 + ||U_t||^2 + ||W_t||^2) \\
     &+ \sum_{t=1}^{T-1}\lambda_2(||U_{t+1}-U_t||^2 + ||W_{t+1}-W_t||^2 \\ 
     &\hspace{0.7cm}+ ||V_{t+1}-V_t||^2),
\end{split}\label{eq:objective}
\end{equation}
where $\lambda_1$ and $\lambda_2$ are regularisation hyperparameters that suppress large weights and large inter-timestep embedding variation, respectively. We refer to this system as \textbf{Cerberus}.

\subsection{Comparison to other systems}\label{sec:model.others}
In our evaluation, we compare Cerberus to the architectures of \citet{yao2018dynamic} and \citet{appel2019temporally} to deconstruct $A$ and/or $C$ as they are defined in Section~\ref{sec:model.source}. As mentioned in Section~\ref{sec:intro}, the former method differs from ours in that it lacks the adjacency source matrix, whereas the latter lacks dynamic word and context embeddings. 
In addition to the architecture differences, our formulations of the $A$ and $C$ matrices also differ subtly from the base systems. 

The original system of \citet{yao2018dynamic} performs a decomposition on a word-word matrix over time to yield dynamic word embeddings, and does not account for users or social structure. It uses the standard matrix factorisation formulation of word embeddings \citep{goldberg2014word2vec} whereby each word is represented according to how often it co-occurs within a small local window with a set of context words. Their premise is that words are similar if they often co-occur with the same words. By contrast, Cerberus uses a social definition of word meaning, and considers two words as being similar if they share similar usage patterns between users. In our evaluations, we use this definition of $C$ in the architecture of \citet{yao2018dynamic} to yield dynamic word and user representations. This model incorporates some social information indirectly since $C$ contains user representations and user similarity is captured through similar language usage, but it does not incorporate the thread structure (who is talking to whom). We refer to this model as \textbf{NoAdj} (\textbf{no} \textbf{adj}acency representation). 

\citet{appel2019temporally} constructs both $A$ and $C$ based on counts rather than PMI. Using PMI instead potentially provides a more discriminative source matrix, since the rarity of words is incorporated. Notably, their system is evaluated in the context of researcher networks, using article keywords to construct the content matrix rather than conversation data, meaning that their data is less sparse. They further use static $V$ and $W$ matrices, but define a similar temporal evolution mechanism in $U$, yielding dynamic user embeddings but static word embeddings. We refer to this model (with source matrices as defined in Section~\ref{sec:model.source}) as \textbf{StatCont} (\textbf{stat}ic \textbf{cont}ent representation).

\subsection{Further optimisations}\label{sec:model.optimisations}
We use the architecture of \citet{appel2019temporally} as basis in this work, and incorporate four further optimisations beyond the architecture changes:
\paragraph{Relaxing the non-negativity constraint} We follow \citet{yao2018dynamic} in relaxing the non-negativity constraint in the matrix factorisation, as the task does not inherently require non-negativity in the learned embeddings.
\paragraph{Downweighing zeroes} Both the adjacency and content matrices are highly sparse, with a sparsity ratio exceeding 0.99. For both source matrices, the data is analogous to an implicit feedback setting, since a zero does not imply a negative score but only signifies a lack of interaction. For this reason, we downweigh the losses originating from empty elements in the source matrices. We use a scaling parameter $c_0=0.01$ to balance the losses originating from the observed and non-observed elements. 
\paragraph{Masking missing users} There is significant user churn between timesteps in the manosphere dataset; as a result, approximately one third of source vectors are empty at each timestep. Since the empty vectors do not tell us anything about the related users, we mask their reconstruction losses. 
\paragraph{Adding biases} Bias terms are used in matrix factorisation applications to capture global trends \citep{koren2009matrix}. For example, some context users in the adjacency matrix may be highly active, meaning that their column might have high values for all users. Interacting with such a context user is not a very informative feature for an individual; as such, a bias term is used to abstract this type of information.

\section{Case study: The manosphere}
\subsection{Background}\label{sec:case}
The models in this study are trained on data from \textit{the manosphere}, which is broadly defined as a collection of communities with a common interest in men’s issues, who are known to engage in highly toxic online behaviour as well as acts of real-world violence \citep{baele2021incel}. These communities have complicated, evolving social structures \citep{ribeiro2021evolution} along with highly specialised and dynamic vocabularies \citep{farrell2019exploring}, which make them well-suited to this study. The roots of the modern manosphere can be traced back to the Men’s Liberation Movement in the 60s and 70s. Since its inception, the movement has fragmented and reformed into several subgroups with related but distinct ideologies \citep{bachaud2024}, also referred to as \textit{categories} in this work. For example, the Men’s Rights Activists (\textbf{MRA}) aim to form a counterpoise to feminist advocacy, while Pickup Artists (\textbf{PuA}) and The Red Pill (\textbf{TRP}) focus on personal development with an aim to seduce women. \textbf{MGTOW} (``Men Going Their Own Way'') promote social separation from women, and \textbf{Incels} (``Involuntarily Celibates'') adopt a nihilistic mindset that promotes self-harm or harm to others. These differing perspectives can cause friction and splintering within the mano\-sphere, as exemplified in the Incels' contempt for the efforts of PuAs to attract partners, and MGTOWs' criticism of the reform efforts of the MRAs \citep{bachaud2024}.

\subsection{Data}
We use the Reddit portion of the manosphere dataset of \citet{ribeiro2021evolution} for training. Any dataset of online conversations with user and thread IDs could be used to train this system; however, the subreddit structure is useful here as it provides labels against which to evaluate the communities discovered by our system (more details in Section~\ref{sec:metrics}). We expect that the system would generalise well to any platform that provides some mechanism for constructing a social graph (for example, a follower network).

The training data consists of posts across 50 subreddits belonging to the 5 abovementioned manosphere categories (annotated by \citet{ribeiro2021evolution}): Incels, MGTOW, PuA, TRP and MRA. It also includes subreddits that are critical of the manosphere (\textit{r/exredpill}, \textit{r/thebluepill}) and subreddits about mental health (\textit{r/depression}, \textit{r/socialanxiety} and \textit{r/suicidewatch}). We retain posts from these subreddits in our training data as they form a useful control mechanism, since they are included in the above dataset and therefore contain the data from the same time period using the same preprocessing, but are expected to be noticeably disparate from the manosphere groups. The number of posts and users per subreddit are shown in Table \ref{tab:subreddit_posts}. 

We are interested in short- to medium-term phenomena and behaviours of communities; as such, we use monthly training windows. We select a training period of 9 months (April--December 2018), consisting of 4,354,116 posts.

As our focus is on evolution over time, we filter the dataset to include only users with interactions across three or more of the included timesteps, resulting in a set of 33,880 users. To construct the adjacency matrix $A$, we further filter the context users (the columns of $A$) to include only the top 10,000 users in terms of post frequency to reduce the computational expense of the matrix factorisation. To construct the content matrix $C$, we include words that are used by more than 20 users; a total of 44,679 words. As a background corpus to calculate the PPMI of the content matrices, we use a dataset of over 40 million Reddit posts by \citet{dziri2019augmenting}.

\subsection{Data limitations}
The \citet{ribeiro2021evolution} dataset was selected because it provides a comprehensive selection of manosphere subcommunities on Reddit, a multi-community platform, which provides proxy labels for our evaluation. This does, however, introduce some limitations. We recognise that extremist communities also gather in platforms that are not readily accessible to outsiders; for example, Telegram and Discord. Given the closed-form nature of these groups, lacking a multi-community structure, it is possible that our method may not generalise well to those groups. However, these groups do sustain a significant presence on Reddit (and other public platforms like Twitter), which provides a recruitment channel for new members. Studying how users relationally evolve in a multicommunity network can provide important insights into the process of radicalisation and radicalisation pathways. Furthermore, these methods are not specific to the study of extremist groups, but can be useful for modelling the evolution of online communities more broadly.

The data used to train our models originates from 2018. While this dataset cannot be expected to capture the latest developments in the manosphere, it provides a comprehensive selection of manosphere subcommunities, and serves to illustrate the utility of our approach. The thematic distinctions between the 5 high-level manosphere groups are not expected to have evolved significantly in the given time period.

\begin{table}[]
    \centering
    \begin{tabular}{|l|l|l|}
    \hline
    \textbf{Category} & \textbf{\# users} & \textbf{\# posts} \\
    \hline
    Incels&9,409&1,700,880\\
    MGTOW&9,797&1,117,866\\
    TRP&8,600&503,842\\
    PUA&3,518&105,970\\
    MRA&6,696&300,073\\
    Mental health&11,788&584,037\\
    Criticism&1333&41,448\\
    \hline
    \end{tabular}
    \caption{Number of users and posts per subreddit in our training data.}
    \label{tab:subreddit_posts}
\end{table}

\section{Evaluation}\label{sec:metrics}
In our evaluation, we aim to isolate the effect of modelling time, social dynamics and language, compared to models that neglected some of these interactions. We therefore follow and expand the evaluation setting of \citet{appel2019temporally}, using two test cases for evaluation: clustering and forecasting. For the latter, we perform an embedding prediction task as well as predicting with which communities a user will interact.

\subsection{Clustering evaluation}
K-means clustering based on the temporal user embeddings $U$ is used to discover sub-communities. Following \citet{appel2019temporally}, these clusters are calculated across different timesteps, meaning that a user may belong to the same cluster or to different clusters over different timesteps. Intuitively, the clusters may be thought of as topics, such that users who do not overlap temporally but have similar embeddings at different timesteps may be assigned to the same cluster. 

Using the subreddit structure, we evaluate the soundness of these clusters based on \textbf{cluster purity}, a widely-used metric for evaluating uniformity of cluster elements as the fraction of users in each cluster that belong to the dominant class \citep{schutze2008introduction}. Since one user may be associated with multiple subreddits at a given timestep, we use a multilabel interpretation of cluster purity, meaning that the dominant class is the class that has the highest representation in a cluster. Let $L(i,t)$ denote the label set of user $i$ at time $t$,  $\Omega = \{\omega_1, \omega_2, \ldots, \omega_K\}$ the set of clusters, and $\mathbb{C} = \{c_1, c_2, \ldots, c_J\}$ the set of classes. We report the average cluster purity over $K$ clusters, given:
\begin{equation}
\text{Purity}(\omega_k, \mathbb{C})= \frac{1}{|\omega_k|} \max_{j} \left| \{ (i,t) \in \omega_k : c_j \in L(i,t) \} \right|.
\end{equation}

Since all users are not present in the dataset at all timesteps, we only evaluate against users for whom there are labels in a given timestep. 

The class labels comprise a two-level hierarchy, consisting of a more general category label (with 7 classes: Incels, MGTOW, TRP, PuA, MRA, mental health and criticism) and a more finegrained subreddit label (with 50 classes). Allowing for some subgroup specialisation at different levels of granularity,
we report results for $k\in\{10,100,1000\}$, using both subreddit and category-level labels. Since the clustering algorithm is dependent on random initialisation, we report the mean and standard deviation over 5 runs. 
\begin{table*}[hbt!]
    \centering
    \begin{tabular}{|l|l|l||l|l||l|l|}
    \hline
        &\multicolumn{2}{c||}{\textbf{K=10}}&\multicolumn{2}{c||}{\textbf{K=100}}&\multicolumn{2}{c|}{\textbf{K=1000}}\\
        \hline
        \textbf{Model} & \textbf{Category} & \textbf{Subreddit} &\textbf{Category} & \textbf{Subreddit}&\textbf{Category} & \textbf{Subreddit}\\
        \hline
        MatFact &$0.36\pm0.001$&$0.33\pm 0.003$&$0.44\pm0.009$&$0.39\pm 0.012$&$0.63\pm0.004$&$0.59\pm0.004$\\
        SharedMF&$0.38\pm0.015$&$0.35\pm0.022$&$0.46\pm0.005$&$0.42\pm0.006$&$0.66\pm0.007$&$0.62\pm0.003$\\
        \hline
        StatCont &$0.55\pm 0.006$ &$0.55\pm0.0134$
        &$0.62\pm 0.006$&$0.55\pm0.005$
        &$\mathbf{0.77\pm0.005}$&$0.71\pm 0.001$ \\
        NoAdj &  $0.59\pm0.022$&$0.50\pm0.005$
        & $\mathbf{0.63\pm0.006}$&$0.57\pm0.008$
        &$0.72\pm0.002$& $0.70\pm0.006$\\
        Cerberus  &$\mathbf{0.61\pm0.025}$& $\mathbf{0.57 \pm 0.050}$ & $\mathbf{0.63\pm0.020}$ & $\mathbf{0.58\pm0.009}$
            & $\mathbf{0.77\pm0.00}$&$\mathbf{0.74\pm0.005}$\\
     \hline
    \end{tabular}
    \caption{Cluster purity at $K=\{10,100,1000\}$, comparing to subreddit and category labels.}
    \label{tab:cluster_purity}
\end{table*}
\subsection{User embedding prediction} 
Given a dataset with $T$ training windows, a user is represented as a sequence of embeddings $\{u_1,u_2,...,u_T\}$. In our case, $T=9$, as we have month-long training windows for 9 months of data. Autoregressive models can be used to forecast an embedding for a user at timestep $T+1$.

We train LSTM-based neural network models to perform this task for each temporal embedding method. Users are split into a train, test and validation set with a \{75:15:10\} ratio. For each user $i$, we generate samples such that $X_{i,t}=\{u_1,..,u_{t-1}\}$ and $y_{i,t}=u_t$ for $t\in[2,T]$. Hyperparameter information is provided in Appendix \ref{sec:hypers.lstm}. To evaluate performance on this task, we report the mean cosine distance between the predicted versus true embeddings as produced by a particular approach. 

\subsection{Community prediction}
A caveat of the user embedding prediction evaluation is that embeddings are only calculated for data on which the factorisation model is trained; as such, data leakage might be a concern. Embeddings predicted beyond the training window could still be informative for analyses in relation to those learned by the model, but for the purposes of evaluation, a direct comparison is not possible. However, we do have access to the true interactions of users in the timestep following the training window. To evaluate the out-of-sample predictive capabilities of the system, we relate the predicted embedding at $t=T+1$, the period immediately following the training window, to the true interactions of a user over the various communities. 

We use a distance-based approach for this purpose. We find the centroid of each community, i.e.\ the mean of the embeddings of all users who interacted with said community in $t_1$ to $t_T$. We then calculate the cosine similarity $s$ between the predicted user embedding and the centroid of each community. The inferred user label is given by the max-normalised similarities to all centroids, i.e. $\hat{y_{i,t}} = \frac{s_{i,t}}{max(s_{i,t})}$, to capture the relative relevance of each community to a user. 

These predictions are compared to the relative engagement volume per community in $t_{T+1}$. Let $f_{i,t}$ denote a vector of interaction counts for a given user over all communities, in a given timestep $t$. Then, the user label is given by $y_{i,t} = \frac{f_{i,t}}{max(f_{i,t})}$.

This formulation equates to a multilabel regression task, which we evaluate using the concordance index (CI; \citealp{harrell1982evaluating}). The CI is the fraction of concordant predictions (i.e. \ ($y_i > y_j$, $\hat{y}_i > \hat{y}_j$) or ($y_i < y_j$, $\hat{y}_i < \hat{y}_j$); in other words, a pairwise comparison is performed on all possible pair permutations, and the fraction that are correctly ordered is calculated. We report both the mean per-class CI (comparing predictions across users for a specific community) and the within-sample CI (evaluating the orderings per user). 

\section{Implementation details}
We extend the TensorFlow implementation of \citet{appel2019temporally} to implement the proposed model. To perform hyperparameter tuning and to establish the efficacy of the four training optimisations outlined in Section \ref{sec:model.optimisations}, we run preliminary experiments on a reduced training set. These results, reported in Appendix \ref{sec:prelim.exp}, show that the optimisations provide statistically significant improvements in the reconstruction accuracy in a smaller scale evaluation setting, and as such they are incorporated in all models that are compared in this work. Our code and models will be released to the community. 

The asymptotic complexity of our method matches that of the approach proposed by \citet{appel2019temporally}. The introduction of temporal word embeddings in our method adds ($T-1$) additional matrix multiplications to the objective function, where $T$ is a constant scalar. Since the number of additional operations is proportional to $T$, the overall asymptotic complexity of our method remains unchanged. The Cerberus system was trained on a Google Cloud instance with 256GB memory and 64CPUs for approximately 1 week. Reducing the training costs would be trivial if required, as the interaction matrices are sparse and do not need to be read into memory in full. The vocabulary could also be reduced by excluding uncommon words or stopwords. Additionally, GPUs can be explored for faster training. In these evaluations, we keep the implementation as-is to replicate the experimental setting used in the precursor works. 

\section{Results}\label{sec:results}
\subsection{Cluster purity}\label{sec:cluster_res}
Cluster purity over $K$ clusters is shown in Table~\ref{tab:cluster_purity}. Our model outperforms the StatCont and NoAdj systems in 4 out of 6 cases, with the largest improvement observed for subreddit purity at $K=1000$. The NoAdj model has the same score as ours for category purity at $K=100$, whereas the StatCont model performs the same as ours for category purity at $K=1000$. We note that cluster purity is better at larger $K$, which is to be expected; purity will be 1 if $K=N$, but the comparison of different models at a given $K$ gives us a useful comparison point. In all cases, the cluster purity is higher for the category labels than the subreddit labels, which again is to be expected; since users are distributed over more classes, clusters are less likely to be dominated by a single subreddit. 

We also compare our cluster purity to embeddings that are obtained through standard static (i.e.\ time-aggregated) matrix factorisation of the content embedding (\textbf{MatFact}) and static shared matrix factorisation of content and adjacency (\textbf{SharedMF}). In this case, only one embedding is produced per user and therefore a user is assigned to only one cluster. We calculate purity based on all communities with which a user interacted over the training period. From the results, we note that the cluster purity is lower in this setting. A possible reason for this result is that users vary in their category associations over time, which is not captured well by the time-aggregated clustering. The static shared matrix factorisation model outperforms the static model without social information. 

The per-cluster purity for $K=10$, at the category level and using Cerberus, is shown in Table~\ref{tab:cat_purity}, sorted by cluster size. Size is based on the number of (\textit{user, timestep}) tuples in the cluster. We note that there is a substantial difference between the high-scoring and low-scoring clusters, with clusters 2, 7, 8, 9 and 10 being high-purity clusters representing 4 out of the 7 higher level categories in the data. With the exception of cluster 2, these are all relatively small clusters, possibly representing the dedicated core users of each ideology who do not engage with other categories. Cluster 2 is a large and high-purity cluster which represents the mental health category. This provides support for the soundness of the clustering and embeddings, as we would expect this control group to be disparate. 

Interestingly, half of the clusters are associated with the MGTOWs, even though only two MGTOW subreddits are included in the dataset; however, these clusters mostly have low purity scores.  Looking at cluster 5, we observe that 20.4\% of its users are MRA contributors, 24.2\% from TRP and 18.4\% from Incels. This indicates that MGTOW users are more likely to have some interaction and/or overlapping interests with other categories of the manosphere, since they form the majority in mixed (low-purity) clusters. The same pattern is observed for larger $K$, with 50 clusters being associated with MGTOW at $K=100$.
\begin{table}[]
    \centering
    \begin{tabular}{|l|l|l|l|}
    \hline
    \textbf{Index}&\textbf{Mode} & \textbf{Purity} &\textbf{Size} \\
    \hline
    1&MGTOW&0.2531&128,256\\
    2&Mental health&0.8728&59,502\\
    3&TRP&0.711&27,104\\
    4&MGTOW&0.3676&26,973\\
    5&MGTOW&0.3439&22,176\\
    6&MGTOW&0.5037&21,260\\
    7&Incels&0.994&10,823\\
    8&MGTOW&0.9555&4,273\\\
    9&TRP&0.8996&2,981\\
    10&MRA&0.8457&1,572\\
    \hline
    \end{tabular}
    \caption{Category purity for $N=10$.}
    \label{tab:cat_purity}
\end{table}
\subsection{Evolution of users} \label{sec:user_evolution}
\begin{figure*}[h!]
    \centering
    \includegraphics[width=\linewidth]{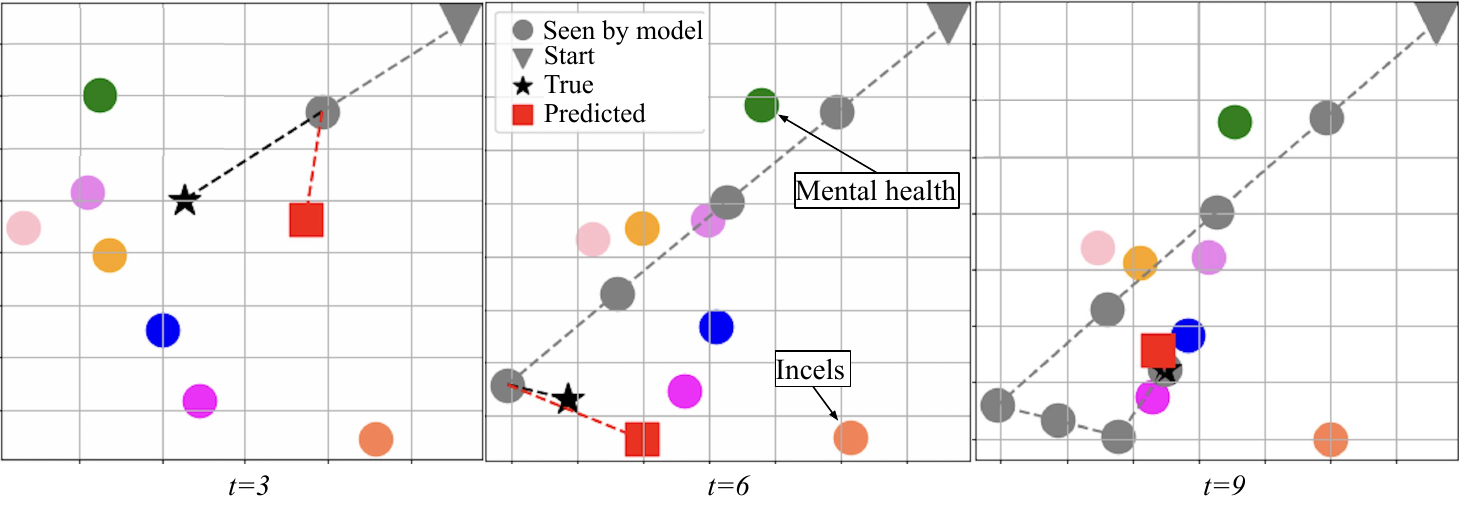}
    \caption{Predicting user evolution over time. Grey markers represent the trajectory of a given user, whereas coloured dots are community centroids. At each timestep, the predicted next embedding is shown in red and the true next embedding in black, while the data seen by the model is in grey.}
    \label{fig:upred}
\end{figure*}
In the context of extremist groups, the ability to represent a user as a dynamic entity at different timesteps is a key benefit of the temporal factorisation approach. Prior work \citep{de2024investigating,ferrara2016predicting} developed predictive models to anticipate whether a user will engage with an extremist group; however, these works relied on derived indicators, whereas we our system models users and communities in the same embedding space. 

Figure \ref{fig:upred} illustrates the embedding prediction task, using PCA to project the embeddings for $t\in\{3,6,9\}$ to 2 dimensions. These predictions can be compared to the centroids of different communities, which also change over time. In this case, we can see that the user starts out as being closest to the mental health community, and moves progressively closer to the manosphere subcommunities.  Our model predicts the right relative direction at each point, and the prediction becomes progressively closer to the true value as it is exposed to more information. After maintaining the same direction of change for the first 5 timesteps,  the model accurately predicts the sharp deviation at $t=6$.

As described in Section \ref{sec:metrics}, we evaluate user evolution modelling in two ways: embedding prediction (comparing learned embeddings to predicted embeddings, within the training window) and community prediction (predicting beyond the training window and comparing to true interactions). These results are shown in Table \ref{tab:emb_pred}. Though we use the subreddit structure to evaluate our models, they can be applied in forums without explicit subcommunities. 

Our model outperforms the antecedent works in the embedding prediction task ($P<0.05$, using the randomised permutation test with Monte Carlo approximation and $N=9999$). Similarly, for the community prediction, our model has the highest CI score for the within-sample and the per-class evaluations. Looking at the per-class breakdown, Cerberus has the highest scores for 5/7 categories. The NoAdj model outperforms the StatCont model for the embedding prediction and the aggregated community prediction evaluations, illustrating that dynamic content representation is of significant importance. 

\setlength{\tabcolsep}{1pt}
\begin{table}[]
    \centering
    \begin{tabular}{|l|p{1.42cm}|p{1.2cm}|p{1.5cm}|}
    \hline  
    \textbf{Evaluation} & \textbf{StatCont} & \textbf{NoAdj} &\textbf{Cerberus}\\
    \hline
    \multicolumn{4}{|c|}{\textbf{Embedding prediction}}\\
    \hline
    Cosine similarity &0.847 &0.876 & \textbf{0.902} \\
    \hline
    \multicolumn{4}{|c|}{\textbf{Community prediction}}\\
    \hline
    CI, within sample &0.702& 0.722&\textbf{0.734} \\
    CI, per class  &0.669&0.674 & \textbf{0.717}\\
    \hspace{1em} MGTOW & 0.695&0.683&\textbf{0.740}\\
    \hspace{1em} TRP & 0.708&0.710&\textbf{0.751}\\
    \hspace{1em} Incels & 0.718&0.634&\textbf{0.764}\\
    \hspace{1em} MRA & \textbf{0.714}&0.708&0.700\\
    \hspace{1em} PUA & 0.573&\textbf{0.643}&0.629\\
    \hspace{1em} Mental health & 0.605&0.769&\textbf{0.782}\\
    \hspace{1em} Criticism &0.628&0.556&\textbf{0.651}\\
    \hline
    \end{tabular}
    \caption{Embedding and community prediction results.}
    \label{tab:emb_pred}
\end{table}

\subsection{Evolution of words}\label{sec:word_evolution}

\begin{figure}
\centering
\begin{subfigure}[b]{0.45\textwidth}
   \includegraphics[width=\linewidth]{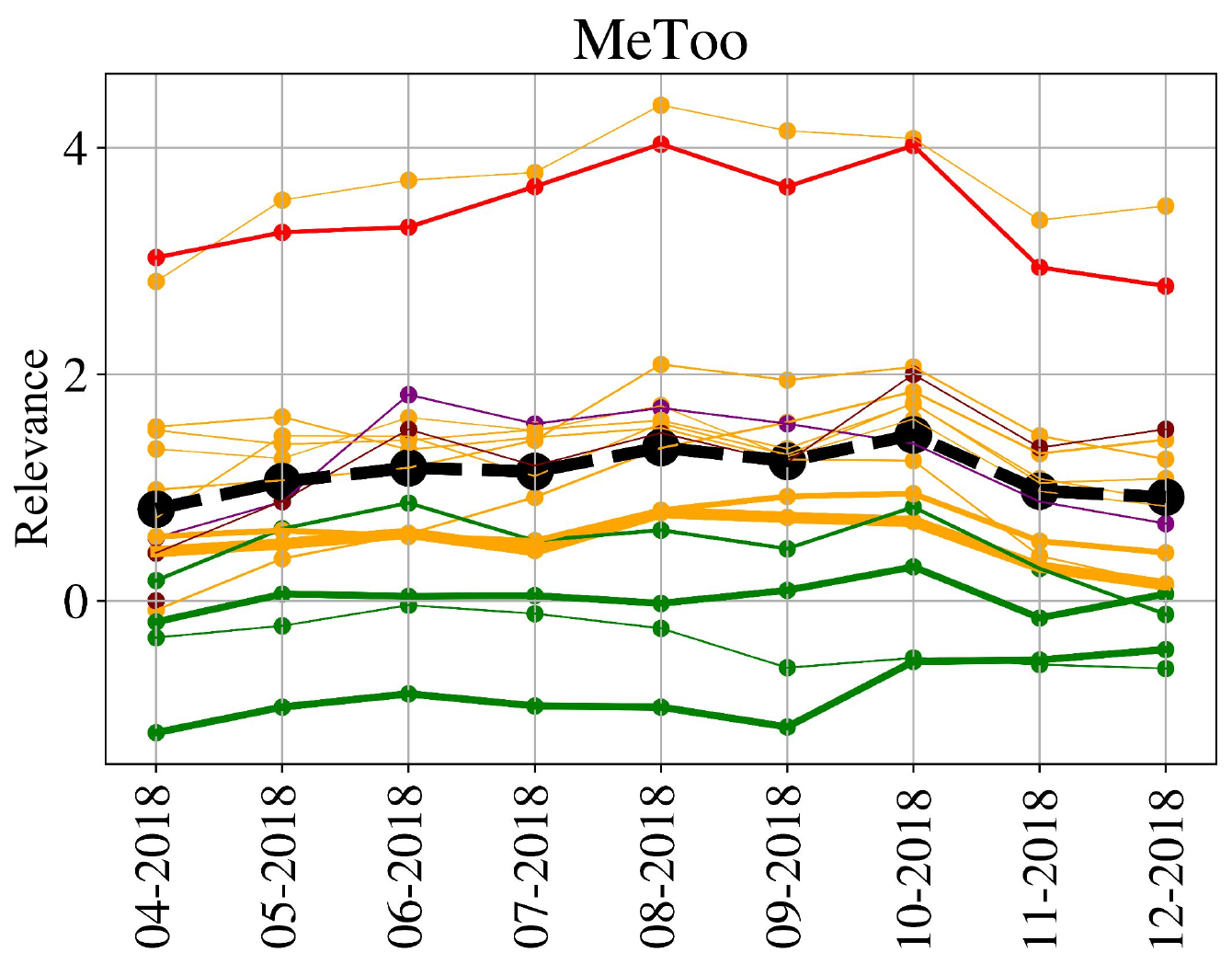}
   \caption{Change in word relevance of "MeToo".}
   \label{fig:metoo}
\end{subfigure}
\begin{subfigure}[b]{0.45\textwidth}
   \includegraphics[width=\linewidth]{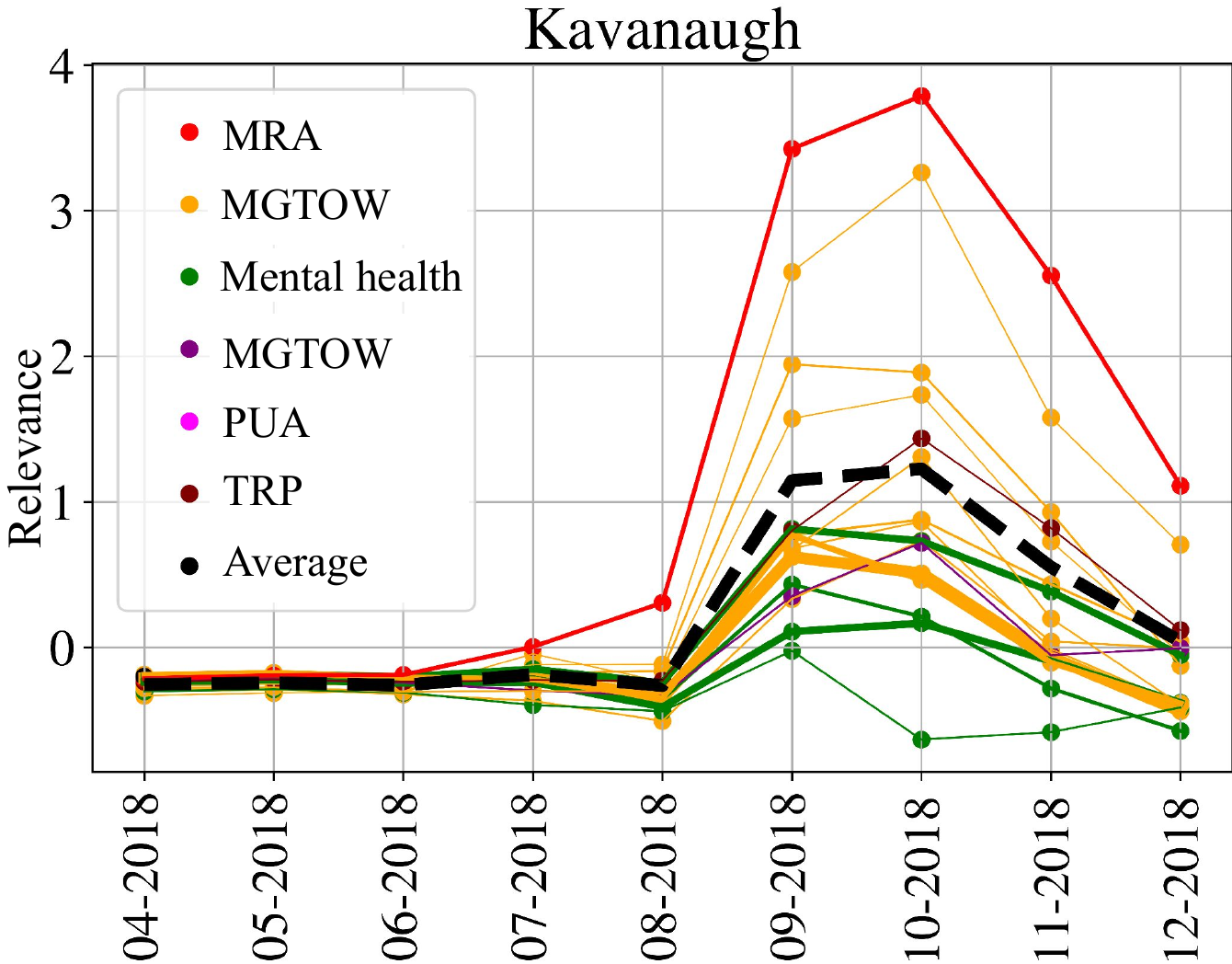}
   \caption{Change in word relevance of "Kavanaugh".}
   \label{fig:kavanaugh}
\end{subfigure}
\caption{Word relevance to different clusters over time.}
\label{fig:word_evolution}
\end{figure}

\begin{figure*}[h!]
    \centering
    \includegraphics[width=0.9\linewidth]{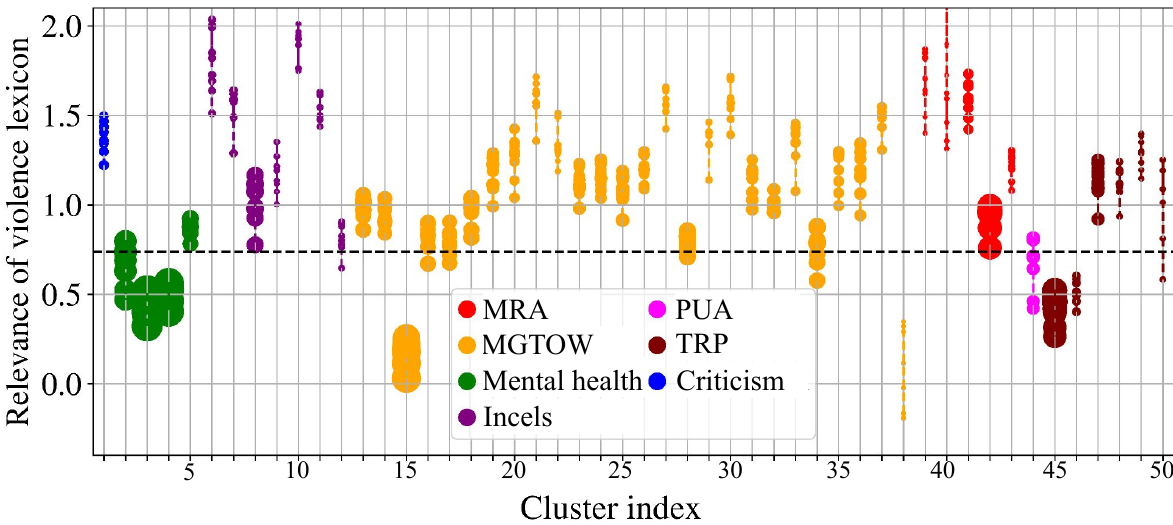}
    \caption{Relevance of the violence lexicon for different clusters.}
    \label{fig:splintering}
\end{figure*}

Capturing the evolution of words is similarly important in the context of extremist groups. This is shown qualitatively in Figure \ref{fig:word_evolution}, using two terms that are tied to temporal events: \textit{MeToo} and \textit{Kavanaugh}. 

We use the clusters as discovered in Section \ref{sec:cluster_res} with $k=100$, and calculate the \textbf{relevance} of a word to a cluster as the dot product of the cluster centroid and the word embedding. Since the clusters are calculated across timesteps, their centroids are fixed, meaning that any changes in word relevance are due to a change in the word embedding. We show only clusters with more than 5 000 users, and scale the linewidth to represent the relative cluster sizes. The cluster labels are determined by the majority class, as per the cluster purity calculation.

For \textit{MeToo}, shown in Figure \ref{fig:metoo}, we note that the word is most relevant to the Men's Rights Activists, who are concerned with legal injustices against men. There is also a small MGTOW cluster for whom it is relevant, but it is of lesser relevance to the larger cluster of MGTOWs and for the mental health forums. The scores are relatively stable over the 9 month training window.  

By comparison, we note a substantial spike in the term \textit{Kavanaugh} during the training window. The term refers to a US justice who was accused of sexual assault. The accusation was widely publicised in September 2018; however, the White House announced in October 2018 that the FBI had found no corroboration of the allegation. Naturally, this event was of great interest to the MRAs, which is reflected in Figure \ref{fig:kavanaugh}\footnote{A neologism \textit{Kavanaughs} was also introduced by the MRAs, to describe men who were falsely accused of rape.}. Given that the community embeddings are fixed, this illustrates that the word embedding changed over time to capture that the term became relevant for specific subgroups in the community, whereas it remained relatively stable for others (e.g.\ the mental health subreddits).

There are three main takeaways from these figures. Firstly, dynamic word embeddings are important: a static model would produce an aggregate representation of \textit{Kavanaugh}, which discards a great deal of information that is likely to be useful for constructing informative user, word and community embeddings. Secondly, there are subgroups within the larger communities who are more and less interested in particular phenomena or events, such as the MeToo movement. This is captured by our model by using individual user representations, rather than treating the movement as monolithic, allowing for the emergence of subgroups with specialised interests. Finally, modelling users and words in the same space provides a strong platform for the analysis of extremist (and other) communities. In the next section, we provide a final exploration of this concept. 

\section{Splintering}\label{sec:splintering}
As discussed in Section \ref{sec:case}, the manosphere shows signs of macro-level splintering, with well-defined subgroups such as Incels forming more violent splinters. Here, we explore lower-level splintering by investigating the clusters produced by our system. For a clearer visualisation, we use K-means clustering with $k=50$. 

To estimate the level of violent language per cluster, we use a subset of 39 terms from the Incel Violent Extremism Dictionary of \citet{baele2023diachronic} that specifically references physical violence (e.g.\ kill, rape, murder) and find the centroid of their Cerberus embeddings. A similar approach for finding a concept centroid by averaging word embeddings is used by \citet{mendelsohn2020framework}. We then find the dot product of each cluster with the violence centroid at every timestep. 
A limitation of this approach is that it does not consider the context in which a word is used (for example, sarcasm); however, by relying on embedding distances of a cluster vector to a concept vector, instead of word counts as used in prior work \citep{baele2023diachronic}, we can obtain a more nuanced measure of the general tendency within the cluster to use language related to violence. While the centroid is constructed based on words that unambiguously relate to physical violence, it should also be close to the representations of emerging slang words that implicitly signal violence within the community. 

The results are shown in Figure~\ref{fig:splintering}. Each vertically-grouped collection represents the violence scores for a specific cluster at different timesteps. As per Section \ref{sec:word_evolution}, the cluster embeddings are fixed, such that variations are caused by changes in word embeddings over time. The marker size corresponds to the cluster size. For reference, the community at index 15 consists of 48,993 user representations. 

For the manosphere communities, we observe a trend of larger, less violent groups and smaller, more violent groups within each category. This supports the existence of low-level splintering; i.e.\ subgroups that are more violent than the mainstream within each category. Future work may investigate the emergence and focuses of these smaller and more violent groups. 

The Incel clusters have the highest mean violence score, which supports the idea that they are a more extreme splinter of the mano\-sphere. The MRAs also show a penchant for violent language, however, this may be due to their interest in discussing sexual assault accusations. This highlights another potential area of future work: accounting for polysemy and linguistic context in the content representation. We observe that the mental health clusters generally have relatively low violence scores, with the exception of one ($i=5$) which has a slightly higher than average score. Upon investigation, we note that this cluster has substantial overlap with the \textit{r/suicidewatch} community. The cluster associated with Pick-up Artists ($i=44$) also has relatively low scores, which reflects their main interest being seduction tricks.

As discussed in Section \ref{sec:cluster_res}, the MGTOW community yields the most subclusters. Future work may investigate this result, as it could be indicative of a more fractured subcommunity structure. We note that the largest MGTOW cluster has a very low violence score, which may be related to the notion that these communities tend to be very supportive and empathetic towards their own members \citep{rich2023}. 

\section{Conclusion}
In this work, we introduce a novel architecture for modelling online communities. Experimental results show that the resulting embeddings yield better results in clustering and embedding forecasting evaluations. Beyond the quantitative improvement, our analyses in Sections \ref{sec:word_evolution} and \ref{sec:splintering} qualitatively illustrate the usefulness of the learned embeddings in the context of an online extremist community. These analyses enable novel insights into the manosphere and its subgroups, and have the potential to support monitoring of these groups. The question of how to act upon such information is less clear and should be treated with caution; past suppression attempts have resulted in splinter groups migrating to specialised platforms. False accusations may have a devastating impact on the lives of those impacted; as such, our approach does not attempt to classify users as ``radicalised'' or ``extremist'' as previous works do, but instead utilises relative distances to communities. Continued cross-disciplinary efforts remain essential in mitigating online harms.

\bibliography{aaai25}

\appendix

\section*{Ethics statement}
The experiments presented in this paper utilise data developed by \citet{ribeiro2021evolution} and shared under a Creative Commons Attribution 4.0 International license. According to \citet{ribeiro2021evolution}, the dataset was created in accordance with the ethical guidelines established by \citet{rivers2014ethical}, which emphasised not attempting to de-anonymise users or connect them across different platforms. Since users interact in a public space on an anonymous basis, there is not a significant risk of infringing on their privacy. Nonetheless, we acknowledge that the users did not consent to have their data analysed for research purposes. Given the genuine and ongoing threats posed by this group, we believe that the use of this data is justifiable. To protect individual identities, we refrain from including any direct quotations or usernames.

We are also conscious of the potential mental health implications for our collaborators and participants. Those who engaged with texts from manosphere communities were encouraged to participate in a support group tailored for researchers dealing with extreme content.

\section*{Paper Checklist}

\begin{enumerate}\itemsep0em

\item For most authors...
\begin{enumerate}
    \item  Would answering this research question advance science without violating social contracts, such as violating privacy norms, perpetuating unfair profiling, exacerbating the socio-economic divide, or implying disrespect to societies or cultures?
    \answerYes{Yes}
  \item Do your main claims in the abstract and introduction accurately reflect the paper's contributions and scope?
    \answerYes{Yes}
   \item Do you clarify how the proposed methodological approach is appropriate for the claims made? 
    \answerYes{Yes; see Section 1 and 2.}
   \item Do you clarify what are possible artifacts in the data used, given population-specific distributions?
    \answerNA{N/A}
  \item Did you describe the limitations of your work?
    \answerYes{Yes; see Sections 5.3, 8 and 9.}
  \item Did you discuss any potential negative societal impacts of your work?
    \answerYes{Yes, see Section 9 and the Ethics statement.}
      \item Did you discuss any potential misuse of your work?
    \answerYes{Yes, see Section 9 and the Ethics statement.}
    \item Did you describe steps taken to prevent or mitigate potential negative outcomes of the research, such as data and model documentation, data anonymization, responsible release, access control, and the reproducibility of findings?
    \answerYes{Yes, see Section 9 and the Ethics statement.}
  \item Have you read the ethics review guidelines and ensured that your paper conforms to them?
    \answerYes{Yes.}
\end{enumerate}

\item Additionally, if your study involves hypotheses testing...
\begin{enumerate}
  \item Did you clearly state the assumptions underlying all theoretical results?
    \answerNA{N/A}
  \item Have you provided justifications for all theoretical results?
    \answerNA{N/A}
  \item Did you discuss competing hypotheses or theories that might challenge or complement your theoretical results?
    \answerNA{N/A}
  \item Have you considered alternative mechanisms or explanations that might account for the same outcomes observed in your study?
    \answerNA{N/A}
  \item Did you address potential biases or limitations in your theoretical framework?
    \answerNA{N/A}
  \item Have you related your theoretical results to the existing literature in social science?
    \answerNA{N/A}
  \item Did you discuss the implications of your theoretical results for policy, practice, or further research in the social science domain?
    \answerNA{N/A}
\end{enumerate}

\item Additionally, if you are including theoretical proofs...
\begin{enumerate}
  \item Did you state the full set of assumptions of all theoretical results?
    \answerNA{N/A}
	\item Did you include complete proofs of all theoretical results?
    \answerNA{N/A}
\end{enumerate}

\item Additionally, if you ran machine learning experiments...
\begin{enumerate}
  \item Did you include the code, data, and instructions needed to reproduce the main experimental results (either in the supplemental material or as a URL)?
    \answerYes{Yes.}
  \item Did you specify all the training details (e.g., data splits, hyperparameters, how they were chosen)?
    \answerYes{Yes; see Appendix A and B, Section 4.2 and Section 6.}
     \item Did you report error bars (e.g., with respect to the random seed after running experiments multiple times)?
    \answerYes{Yes, for metrics where randomness plays a role; see Table 2. We also report statistical significance using a randomised permutation test.}
	\item Did you include the total amount of compute and the type of resources used (e.g., type of GPUs, internal cluster, or cloud provider)?
    \answerYes{Yes; see Section 6.}
     \item Do you justify how the proposed evaluation is sufficient and appropriate to the claims made? 
    \answerYes{Yes; see Section 5.}
     \item Do you discuss what is ``the cost`` of misclassification and fault (in)tolerance?
    \answerNo{We do not perform classifications beyond predicting which community a user will interact with, which does not have a significant misclassification cost. We do discuss the implications of false accusations in Section 9.}
  
\end{enumerate}

\item Additionally, if you are using existing assets (e.g., code, data, models) or curating/releasing new assets, \textbf{without compromising anonymity}...
\begin{enumerate}
  \item If your work uses existing assets, did you cite the creators?
    \answerYes{Yes; section 4.}
  \item Did you mention the license of the assets?
    \answerYes{Yes; see the Ethics statement.}
  \item Did you include any new assets in the supplemental material or as a URL?
    \answerYes{Our code will be shared.}
  \item Did you discuss whether and how consent was obtained from people whose data you're using/curating?
    \answerYes{Yes; see the Ethics statement.}
  \item Did you discuss whether the data you are using/curating contains personally identifiable information or offensive content?
    \answerYes{Yes; see the Ethics statement.}
\item If you are curating or releasing new datasets, did you discuss how you intend to make your datasets FAIR?
\answerNA{N/A}
\item If you are curating or releasing new datasets, did you create a Datasheet for the Dataset? 
\answerNA{N/A}
\end{enumerate}

\item Additionally, if you used crowdsourcing or conducted research with human subjects, \textbf{without compromising anonymity}...
\begin{enumerate}
  \item Did you include the full text of instructions given to participants and screenshots?
    \answerNA{N/A}
  \item Did you describe any potential participant risks, with mentions of Institutional Review Board (IRB) approvals?
    \answerNA{N/A}
  \item Did you include the estimated hourly wage paid to participants and the total amount spent on participant compensation?
    \answerNA{N/A}
   \item Did you discuss how data is stored, shared, and deidentified?
   \answerNA{N/A}
\end{enumerate}

\end{enumerate}

\section{Preliminary experiments}\label{sec:prelim.exp}
\newcolumntype{P}[1]{>{\centering\arraybackslash}p{#1}}
\begin{table}[h]
    \centering
    \begin{tabular}{|l|P{1.45cm}P{1.4cm}P{1.4cm}|}
    \hline
        & \multicolumn{3}{c||}{\textbf{Content reconstruction error}} \\
        \hline
         \textbf{Model} & \textbf{>0-MAE} & \textbf{0-MAE} & \textbf{WMAE} \\
         \hline
         Chimera &2.650&2.551&2.675\\
         \hline
         \hspace{0.5em} - nonneg. constraint&2.607&\textbf{1.573}&2.623\\
         \hspace{0.5em} + scaling zeroes&2.365&1.742&2.383\\
         \hspace{0.5em} + missing user masking&\textbf{2.332}&2.113&\textbf{2.353}\\
        \hspace{0.5em} + biases &2.350&2.225&2.372\\
        &  \multicolumn{3}{c|}{\textbf{Adjacency reconstruction error}}\\
        \hline
         \textbf{Model} & \textbf{>0-MAE} & \textbf{0-MAE} & \textbf{WMAE}\\
         \hline
         Chimera &2.549&2.543&2.574\\
         \hline
         \hspace{0.5em} - nonneg. constraint&\textbf{1.566}&\textbf{1.566}&\textbf{1.582}\\
         \hspace{0.5em} + scaling zeroes&1.723&1.591&1.739\\
         \hspace{0.5em} + missing user masking&1.703&1.693&1.720\\
        \hspace{0.5em} + biases &1.693&1.795&1.711\\
    \hline
    \end{tabular}
    \caption{Results for the preliminary experiments.}
    \label{tab:rec_prelim}
\end{table}
To reduce the computational overhead of the evaluation, we run preliminary experiments to establish the efficacy of the optimisations outlined in Section \ref{sec:model.optimisations} and to perform hyperparameter tuning. We use a reduced training set of 2 months of data and train for 100 epochs per system. A validation set is constructed by randomly masking 10\% of each users' interactions per timestep. We use the mean absolute error (MAE) as metric, which is also used in the objective function in training. Given the sparsity of the dataset, we split the results for nonzero elements (referred to as \textbf{NZ-MAE}) and empty elements (\textbf{0-MAE}), and further report a weighted MAE (\textbf{WMAE}), with a weighting based on the scaling parameter $c_0=0.01$.

\paragraph{Hyperparameter optimisation} We first run a grid search with $\lambda_1,\lambda_2 \in \{0.01,0.1,1\}$ using the basic StatCont model. The best WMAE is recorded for $\lambda_1=0.1$ and $\lambda_2=1$, and we use this in all further experiments. We further use Adam optimisation with $\alpha=0.01$ and a latent factor size of $d=100$.

\paragraph{Training optimisations} We use the StatCont model as basis and apply each of the optimisations sequentially. Statistical significance is measured using the two-sided T-test with $\alpha=0.05$, using the per-user WMAE between consecutive changes. 

Results are shown in Table \ref{tab:rec_prelim}. For the content reconstruction, all changes except for adding biases result in a statistically significant improvement in the nonzero MAE, whereas all but the downweighing of zeroes result in improvements in the adjacency reconstruction. The model with all four changes applied outperforms the base model for both the content and the adjacency reconstruction. We therefore adopt these changes in all further experiments in this paper.

\section{Community prediction model}\label{sec:hypers.lstm}
The community prediction model comprises one LSTM layer with 256 units, followed by two dense layers of 512 and 256 units, respectively. We use a grid search to determine the best dropout and learning rate values for each model, experimenting with $\eta\in\{0.001,0.01,0.1\}$ and $p\in\{0.1,0.2,0.5\}$. The mean squared error is used as training loss, and Adam \citep{kingma2014adam} is used for optimisation.
\end{document}